\UseRawInputEncoding
\documentclass[pdflatex,sn-mathphys]{sn-jnl}
\usepackage{siunitx}
\usepackage{braket}
\DeclareSIUnit{\molar}{M}
\DeclareSIUnit{\Gauss}{G}
\DeclareSIUnit{\Tesla}{T}
\DeclareSIUnit{\Kelvin}{K}
\jyear{2022}%

\theoremstyle{thmstyleone}%
\theoremstyle{thmstyletwo}%

\theoremstyle{thmstylethree}%

\begin{document}

\title[Extending the coherence time of spin defects in hBN]{Extending the coherence time of spin defects in hBN enables advanced qubit control and quantum sensing}

\author*[1,4]{\fnm{Roberto} \sur{Rizzato}}\email{roberto.rizzato@tum.de}
\equalcont{These authors contributed equally to this work.}
\author[3,4]{\fnm{Martin} \sur{Schalk}}
\equalcont{These authors contributed equally to this work.}
\author[1]{\fnm{Stephan} \sur{Mohr}}
\author[1,2]{\fnm{Joachim P.} \sur{Leibold}}
\author[1,4]{\fnm{Jens C.} \sur{Hermann}}
\author[1]{\fnm{Fleming} \sur{Bruckmaier}}
\author[3]{\fnm{Peirui} \sur{Ji}}
\author[5]{\fnm{Georgy V.} \sur{Astakhov}}
\author[5]{\fnm{Ulrich} \sur{Kentsch}}
\author[5]{\fnm{Manfred} \sur{Helm}}
\author[3,4]{\fnm{Andreas V.} \sur{Stier}}
\author[3,4]{\fnm{Jonathan J.} \sur{Finley}}
\author*[1,4]{\fnm{Dominik B.} \sur{Bucher}}\email{dominik.bucher@tum.de}

\affil[1]{\orgdiv{Technical University of Munich, TUM School of Natural Sciences}, \orgname{Department of Chemistry}, \orgaddress{\street{Lichtenbergstra{\ss}e 4}, \city{Garching bei M{\"u}nchen}, \postcode{85748}, \country{Germany}}}

\affil[2]{\orgdiv{Technical University of Munich, TUM School of Natural Sciences}, \orgname{Department of Physics}, \orgaddress{\street{James-Franck-Str. 1}, \city{Garching bei M{\"u}nchen}, \postcode{85748}, \country{Germany}}}

\affil[3]{\orgdiv{Walter Schottky Institute, TUM School of Natural Sciences}, \orgaddress{\street{Am Coulombwall 4}, \city{Garching bei M{\"u}nchen}, \postcode{85748}, \country{Germany}}}

\affil[4]{\orgdiv{Munich Center for Quantum Science and Technology (MCQST)}, \orgaddress{\street{Schellingstr. 4}, \city{M{\"u}nchen}, \postcode{D-80799}, \country{Germany}}}

\affil[5]{\orgdiv{Helmholtz-Zentrum Dresden-Rossendorf}, \orgname{Institute of Ion Beam Physics and Materials Research}, \orgaddress{\street{Bautzner Landstra{\ss}e 400}, \city{Dresden}, \postcode{01328}, \country{Germany}}}


\abstract{
Spin defects in hexagonal Boron Nitride (hBN) attract increasing interest for quantum technology since they represent optically-addressable qubits in a van der Waals material. In particular, negatively-charged boron vacancy centers (${V_B}^-$) in hBN have shown promise as sensors of temperature, pressure, and static magnetic fields. However, the short spin coherence time of this defect currently limits its scope for quantum technology. 
Here, we apply dynamical decoupling techniques to suppress magnetic noise and extend the spin coherence time by nearly two orders of magnitude, approaching the fundamental $T_1$ relaxation limit. Based on this improvement, we demonstrate advanced spin control and a set of quantum sensing protocols to detect electromagnetic signals in the MHz range with sub-Hz resolution. This work lays the foundation for nanoscale sensing using spin defects in an exfoliable material and opens a promising path to quantum sensors and quantum networks integrated into ultra-thin structures.}



\maketitle
\section{Main}\label{sec1}
Optically addressable spin defects in semiconductors are promising systems for various applications in quantum science and technology, including sensing and metrology\cite{weber_2010,awschalom_quantum_2013,degen_quantum_2017,awschalom_2021}. In contrast to other defects typically hosted in 3D crystals\cite{Heremans_2016}, the recently discovered boron vacancy center ($V_B^{-}$) in hexagonal boron nitride (hBN)\cite{abdi_color_2018,gottscholl_initialization_2020} is embedded in a van der Waals material which can be exfoliated down to the limit of a single monolayer\cite{geim_van_2013,novoselov_2d_2016}(Figure \ref{Fig1}a). 
Such a unique feature would be advantageous for a wide range of applications where a minimal spatial separation of the defect to a specific target is highly desired. For example, in nanoscale quantum sensing spatial resolution is determined by the proximity of the defect to the test object\cite{rugar_2013,barry_sensitivity_2020,tetienne_prospects_2021}, or for integrated quantum photonic devices van der Waals materials can be readily exfoliated onto different substrates and used as spin-photon interfaces\cite{awschalom_2021}. Furthermore, the $V_B^{-}$ center, incorporated in ultra-thin hBN foils, allows for easier integration of manipulable qubits in 2D heterostructures. 
This possibility opens up unexplored paths for investigating novel composite materials and phenomena in nanoelectronics, nanophotonics, and spintronics\cite{novoselov_2d_2016,geim_van_2013,miao_2019,lemme_2d_2022,duan_2022}.
The first protocols for generating $V_B^{-}$  centers in hBN have been recently presented\cite{kianinia_generation_2020,gao_femtosecond_2021}, and their spectroscopic characterization has been accomplished in several studies\cite{gottscholl_initialization_2020,gottscholl_room_2021,luxmoore_2022,murzakhanov_electronnuclear_2022,yu_excited-state_2022}. Importantly, by detecting the changes in their optically detected magnetic resonance (ODMR) spectra, the $V_B^{-}$ centers demonstrated to work as sensors for temperature, pressure, and static magnetic fields, in some cases being competitive with more mature spin defect-based sensors\cite{gottscholl_spin_2021}. 
Based on these results, first applications have been demonstrated for the magnetic and temperature nanoscale imaging of low-dimensional materials\cite{healey_quantum_2022, huang_wide_2022}. 

Coherent control of ensembles of $V_B^{-}$ centers has recently been
shown\cite{gottscholl_room_2021,gao_high-contrast_2021,gao_nuclear_2022,haykal_decoherence_2022,luxmoore_2022}, although coherence times on the order of $\sim100$ nanoseconds have been reported\cite{haykal_decoherence_2022,liu_coherent_2022}. These short timescales significantly restrict the utility of such spin qubits and discourage the development of applications based on coherent spin manipulation.

In this work, we overcome this limitation and achieve an over 80-fold extension of the room-temperature $T_2$ coherence time of $V_B^{-}$ ensembles in hBN. The combination of efficient microwave (MW) delivery and precise MW control allowed us to perform dynamical decoupling schemes, such as Carr-Purcell-Meiboom-Gill\cite{meiboom_modified_1958} (CPMG) to efficiently suppress magnetic noise from the spin bath and increase the $V_B^{-}$ coherence.
Furthermore, we generate $V_B^{-}$ dressed-states (DS) by applying  spinlock pulses and show that spin coherence can be preserved for a time ($T_{1\rho}$), which is on the same order as the spin-lattice relaxation time $T_{1}$ and nearly 2 orders of magnitude longer than the spin-echo $T_2$ time.
These improvements enable the detection of radiofrequency (RF) signals with a frequency resolution far beyond the intrinsic spin defect coherence time. This work broadens the applicability of $V_B^{-}$ defects in hBN, opening up new opportunities for nanoscale quantum technologies.  
\section{Characterization of $V_B^{-}$spin properties}\label{sec2}
\begin{figure}[htbp]%
\centering
\includegraphics[width=1\textwidth]{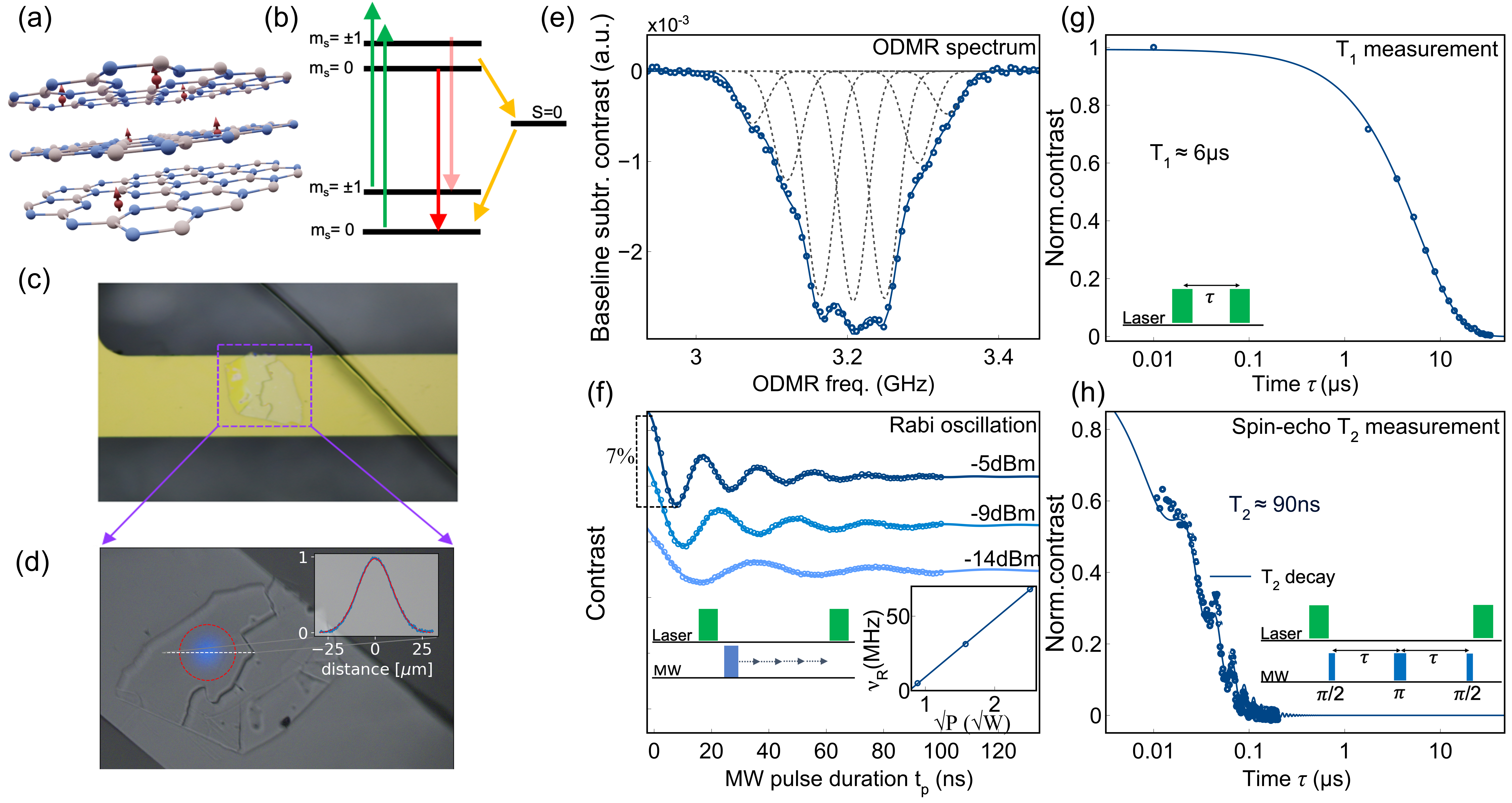}
\caption{\textbf{Characterization of the $V_B^{-}$ defect's spin properties.}
a) Idealized structure of $V_B^{-}$ spin defects (red arrows) in hBN layers. Nitrogen atoms are displayed in blue, boron in grey. The number of the spin defects is only for illustrative purpose and does not reflect the actual density.
b) Simplified energy levels displaying the ground and excited triplet states, the optical excitation transitions (green arrows), the fluorescence pathways (red arrows) and the non-radiative pathways through ISC to the S=0 singlet state (yellow arrows) which allows for spin-state dependent optical detection. 
c) Microscope image of the hBN flake positioned on the gold MW stripline. 
d) Image of the sample from the ODMR setup with the laser spot (blue, false color) for the initialization/interrogation of the spin ensembles. In the inset the Gaussian profile of the laser spot is shown. 
e) ODMR spectrum of the $\vert0\rangle\rightarrow\vert-1\rangle$ transition. The data points are fitted with single Gaussian functions (dotted lines) that are summed up to give the overall spectral lineshape (solid-blue line).
f) Rabi oscillations for four different MW amplitudes. Inset, left: pulse sequence for the Rabi nutation experiment. Right: dependence of the Rabi frequency $\nu_R$versus the square root of microwave power P.
g) Semi-log plot of the spin-lattice relaxation time decay ($T_1$) measured by the recovery pulse sequence (inset). A $T_1$ time of $\sim\SI{6}{\micro\second}$ is extracted from the mono-exponential fit.
h) Semi-log plot of the coherence time $T_2$ measured with the depicted spin-echo sequence. $T_2=\SI{90}{\nano\second}$ is obtained from a stretched exponential fit. A strong signal modulation is superimposed to the same $T_2$ decay curve with a frequency of $\sim\SI{45}{\mega\hertz}$.}\label{Fig1}
\end{figure}

The negatively-charged boron vacancy center (${V_B}^-$) consists of a missing boron atom in the hBN lattice surrounded by three equivalent nitrogen atoms (Figure \ref{Fig1}a). 
Ten electrons occupying six defect orbitals result in a triplet $S=1$ ground state consisting of the $\vert m_{s}=0\rangle$ ($\vert 0\rangle$) and $\vert m_{s}=\pm1\rangle$ ($\vert \pm1\rangle$) spin-states\cite{abdi_color_2018,ivady_ab_2020}. At zero magnetic field, the $\vert \pm1\rangle$ states are degenerate but separated in energy from the $\vert 0\rangle$ state due to a zero-field splitting (ZFS) of $D\sim\SI{3.47}{\giga\hertz}$. The transition from the ground state to the excited state by green laser illumination (e.g., $\lambda=\SI{532}{\nano\meter}$) is followed by a phonon-assisted radiative decay with broad photoluminescence (PL) peaking at $\sim\SI{850}{\nano\meter}$\cite{qian_unveiling,qian_2022}. A spin-state-dependent relaxation path through inter-system crossing (ISC) leads to two important consequences: 1) the ${V_B}^-$ defects can be optically initialized into the $\vert0\rangle$ state under ambient conditions; 2) the $\vert0\rangle$ and $\vert \pm1\rangle$ states can be distinguished by their spin-state dependent PL (Figure  \ref{Fig1}b)\cite{reimers_2020,ivady_ab_2020,gottscholl_initialization_2020,luxmoore_2022}. 

All experiments presented in this work were conducted under ambient conditions on $V_B^-$ ensembles obtained by He\textsuperscript{+} implantation of hBN flakes ($\sim\SI{100}{\nano\meter}$-thick, details in the Methods section). As depicted schematically in Figure \ref{Fig1}c, the flakes are directly transferred onto a gold MW microstripline that is used for $V_B^-$ spin manipulation. A microscope for spatially resolved ODMR measurements has been built that can address defined areas of the sample with a laser spot size of $\sim\SI{20}{\micro\meter}$ diameter (see Figure \ref{Fig1}d and Experimental setup in Methods).

As a first characterization experiment, we perform ODMR at a bias magnetic field $B_0\sim\SI{8}{\milli\Tesla}$. The spectrum shows a broad resonance corresponding to the $\vert  0 \rangle \rightarrow \vert -1 \rangle$ transition (Figure \ref{Fig1}e), revealing characteristic features due to the strong hyperfine (HF) coupling of the $V_B^{-}$ electronic spin with the three equivalent \textsuperscript{14}N nuclei. The HF lines can be fitted with seven Gaussian functions, separated by $\sim\SI{44}{\mega\hertz}$\cite{gottscholl_initialization_2020,gottscholl_room_2021,haykal_decoherence_2022,gao_nuclear_2022}. Their lineshape indicate the inhomogeneously broadened nature of the spectrum, where the electronic spins experience a broad distribution of local magnetic fields due to the intricate HF structure. From this first observation, one can expect
that fast decoherence might dramatically affect the $V_B^{-}$ electronic spins. 
To perform coherent control of the ${V_B}^-$ centers, we run Rabi experiments with the MW frequency at the central peak of the ODMR spectrum and monitor the fluorescence contrast while sweeping the MW pulse duration $t_p$. Figure \ref{Fig1}f shows Rabi oscillations for different MW amplitudes. We observe a $6-7\%$ fluorescence contrast at maximum amplitude with a $\pi$-pulse duration of $t_p^{(\pi)}=\SI{7.5}{\nano\second}$, corresponding to a Rabi frequency of $\nu_R\sim\SI{67}{\mega\hertz}$. The maximization of the Rabi frequency is crucial for two reasons: 1) it allows for short pulses that are necessary for an efficient spin manipulation, especially in the presence of fast spin-dephasing, 2) it allows us to drive a large bandwidth of the ODMR spectrum. Fourier transformation of our rectangular $\pi/2$-pulses would give an excitation bandwidth of approximately $1/t_p^{(\pi/2)}=\SI{250}{\mega\hertz}$ for a $\sim\SI{4}{\nano\second}$ duration. This has the positive effect of increasing the observable contrast whilst reducing detrimental spectral diffusion effects.  

To characterize the $V_B^{-}$ relaxation properties, we measure the spin-lattice relaxation time $T_1$ using a protocol consisting of two $\SI{5}{\micro\second}$-long laser pulses for initialization and readout separated by the sweep time $\tau$. A time constant $T_1\sim\SI{6}{\micro\second}$ is extracted from a mono-exponential fit of the resulting contrast decay (see Figure \ref{Fig1}g). 
We continue to measure the native coherence time $T_2$ using the spin-echo sequence depicted in Figure \ref{Fig1}h. Here, the fluorescence contrast is detected while sweeping the free precession time $\tau$. By fitting the resulting signal decay with a stretched exponential function\cite{bar-gill_suppression_2012,bar-gill_solid-state_2013} (see Pulse sequences, normalizations and fittings in Methods), a time constant $T_2\sim\SI{90}{\nano\second}$ is extracted, consistent with recent literature\cite{haykal_decoherence_2022,liu_coherent_2022}. Such a coherence time is short, especially compared to other ensembles of spin defects in 3D-host materials, where typical $T_2$ times lie in the vicinity of the microsecond timescales at room temperature\cite{barry_sensitivity_2020}.
Interestingly, a clear oscillation appears superimposed on the spin-echo decay. This modulation is tentatively assigned to the interaction of the $V_B^-$ electronic spin with the three \textsuperscript{14}N nuclei since the frequency of $\sim\SI{45}{\mega\hertz}$, fit by a cosine function (see Methods and Supplementary Note 1), matches the HF coupling\cite{gottscholl_initialization_2020}. 
An improvement of the coherence time is a critical step for utilizing $V_B^{-}$ in hBN in advanced quantum technologies, where qubit coherent manipulation and quantum information storage/retrieval are each essential preconditions.
\section{Extension of the $V_B^{-}$ coherence}\label{sec3}
Dynamical decoupling (DD) techniques are traditional tools of nuclear magnetic resonance spectroscopy\cite{Carr_Purcell_1954,meiboom_modified_1958} and have been extensively applied in the past years to prolong the coherence of spin defects in solid-state materials\cite{ryan_robust_2010,de_lange_universal_2010,naydenov_dynamical_2011,pham_enhanced_2012,abobeih_one-second_2018,barry_sensitivity_2020,anderson_five-second_nodate}. 
Here, we apply this approach to $V_B^{-}$ ensembles in hBN to improve their short $T_2$ times and unlock possibilities based on advanced spin manipulation. 

\begin{figure}[htbp]%
\centering
\includegraphics[width=0.85\textwidth]{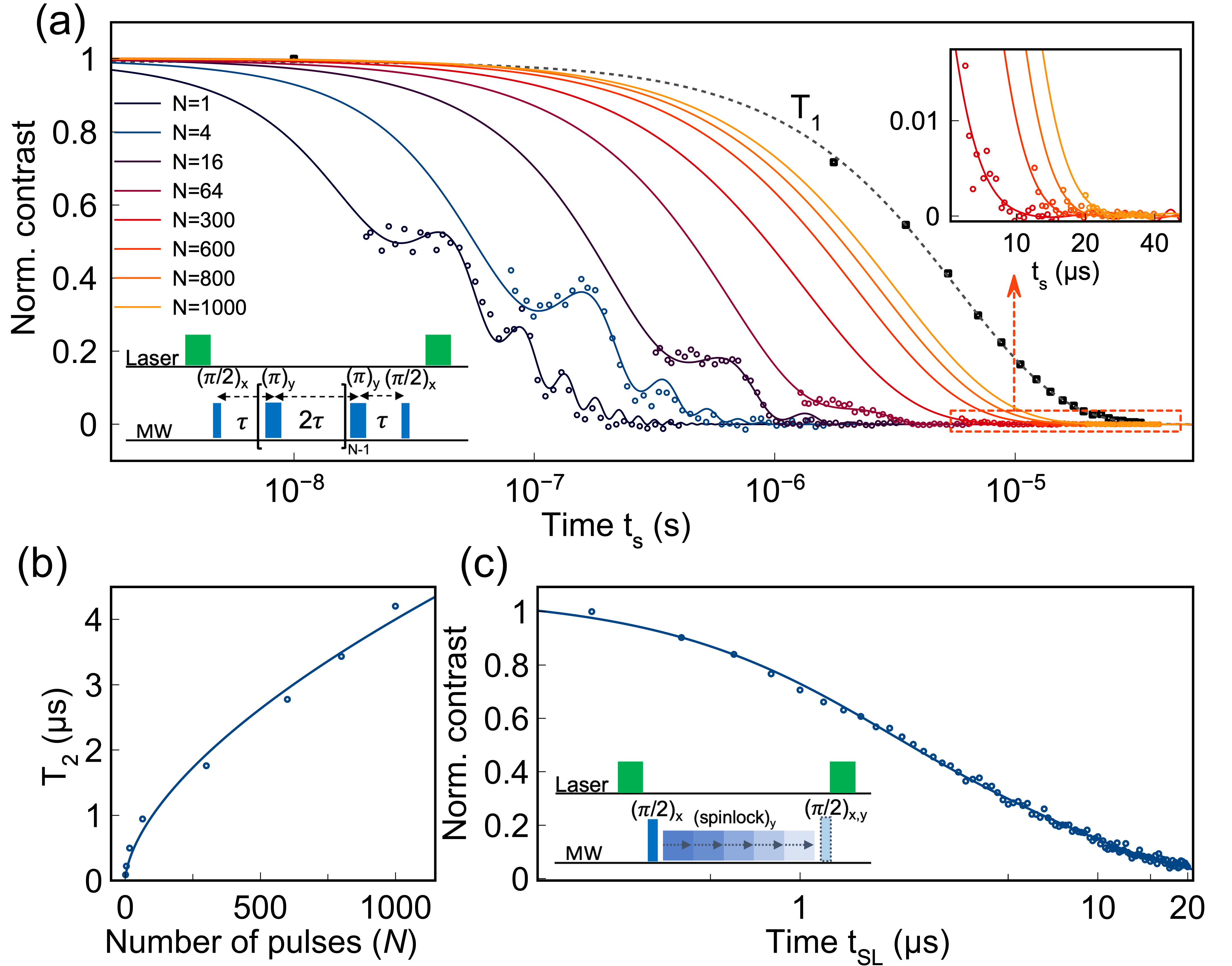}
\caption{\textbf{Extension of coherence for $V_B^{-}$ in hBN.} 
a) Measured decoherence curves (semi-log plot) for the $V_B^-$ spins under the effect of different CPMG decoupling pulse sequences with increasing number of $\pi$-pulses \textit{N}. Coloured dots and solid lines are data points and fits, respectively. In black, the $T_1$ measurement of Figure \ref{Fig1}g is shown with the corresponding fit as dotted line. Bottom, left: Dynamical decoupling (CPMG) pulse sequence. Top right: enlarged plot area for better visualization of the data points in the dotted red rectangle.
b) Plot of the $T_2$ values extracted from fitting the curves in (a) vs the number of $\pi$-pulses $N$ utilized in each CPMG experiment.
c) Rotating frame spin-lattice relaxation time ($T_{1\rho}$) measured with the depicted spinlock sequence. 
}\label{Fig2}
\end{figure}

A significant source of $V_B^{-}$ decoherence is likely to be found in spin `flip-flops' from the bath spins (nuclei or paramagnetic impurities) surrounding the defects\cite{haykal_decoherence_2022,liu_coherent_2022}. These processes cause random  magnetic field fluctuations felt by the electronic spin on a time scale set by the average interactions involved\cite{sousa2009electron,pham_enhanced_2012,bar-gill_suppression_2012,ye_spin_2019}. We show that the CPMG dynamical decoupling protocol\cite{Carr_Purcell_1954} can be applied to decouple the $V_B^{-}$ spins from magnetic noise. This is done by applying resonant MW $\pi$-pulses, following the scheme $(\pi/2)_x[-\tau-(\pi)_y-\tau]_N$ (see inset in Figure \ref{Fig2}a), that have the effect to periodically re-phase the $V_B^{-}$ superpositions and sustain their coherence over longer timescales. Then, the coherence is mapped to the spin populations via a last $(\pi/2)_x$ pulse for the final optical readout. If $\tau$ is shorter than the spin bath fluctuation correlation time $t_c$, the magnetic noise  would appear to be time-independent\cite{sousa2009electron, bar-gill_suppression_2012}, and the train of $\pi$-pulses can effectively cancel it out.

In Figure \ref{Fig2}a, we show the signal intensity of the spin-echo obtained after $N$ MW $\pi$-pulses upon increasing the delay $\tau$ between them. We plot the decays against the total pulse sequence time $t_s=2N\tau$. This results in multiple decay curves, showing how fast decoherence occurs depending on the number of $\pi$-pulses utilized for noise suppression. Then, we extract the characteristic $T_2$ times by fitting every curve (see Methods). We observe a factor $\sim50$ increase in the coherence time ($T_2^{(1000)}\sim\SI{4.2}{\micro\second}$) by applying up to 1000 $\pi$-pulses, with respect to the spin-echo with a single $\pi$-pulse ($T_2\sim\SI{90}{\nano\second}$). 
The enhanced coherence times closely approach the spin-lattice relaxation time, which is the theoretical limit\cite{slichter2013principles}. 
We note that dynamical decoupling does not affect the strong modulation in the spin-echo $T_2$ measurement. We also note that the finite duration of the MW pulses sets the shortest pulse sequence time $t_s$. Moreover, particularly for large $N$, we expect contributions from pulse errors and $T_1$ relaxation\cite{naydenov_dynamical_2011,pham_enhanced_2012}.

In Figure \ref{Fig2}b, we show the increase of the $T_2$ times versus the number of $\pi$-pulses $N$. The plot depicts a sub-linear dependence that can be fit with a simple power function $f(N)=a\times N^s$ where $s\sim0.60$, which is in good agreement with the theoretical dependence of $T_2\propto N^{2/3}$ expected for a Lorentzian spin bath, in the limit of long spin bath correlation times $t_c\gg \tau$\cite{sousa2009electron,bar-gill_suppression_2012}.
These results demonstrate that dynamical decoupling works effectively for our sample, similar to the situation encountered for nitrogen-vacancy(NV) ensembles in diamond in the presence of high-density paramagnetic impurities (50-100 ppm) and natural abundance \textsuperscript{13}C nuclear spins\cite{bar-gill_suppression_2012,pham_enhanced_2012} or for $V$\textsubscript{Si}$^-$ in 4H-SiC\cite{simin_2017}.

We also note that a factor of two increase in coherence time by a CPMG protocol was shown by conventional electron paramagnetic resonance (EPR) spectroscopy, although under very different experimental conditions (magnetic field of $\sim \SI{3}{\Tesla}$ and cryogenic temperature of $\sim\SI{50}{\Kelvin}$)\cite{murzakhanov_electronnuclear_2022}.

An alternative approach that preserves coherence relies on generating $V_B^{-}$ dressed spin states using spinlock pulse sequences\cite{schweiger_principles_2001,london_detecting_2013}. After optical initialization of the defects in the $\vert$0$\rangle$ state, a $(\pi/2)_x$ pulse generates their coherent superposition which is then locked along the $y$-axis of the Bloch sphere by a spinlock pulse. We measure the spin-lattice relaxation time in the electron-spin rotating frame ($T_{1\rho}$) with the experiment depicted in Figure \ref{Fig2}c. 
Here, the $V_B^{-}$ fluorescence contrast is monitored while increasing the spinlock pulse duration. We demonstrate that the $V_B^{-}$ dressed-states and, therefore, the electronic spin coherence can survive $\sim$80x longer than in the case of the spin-echo $T_2$ (Figure 1h). The decay of the spinlocked coherence can be fit by a bi-exponential function (see Methods) and reaches the $\sim T_1$ relaxation time.

\section{Sensing RF signals with $V_B^{-}$ defects in hBN}\label{sec4}
With the improved coherence time, the application of advanced quantum sensing protocols is now possible. 
In particular, we explore oscillating (AC) magnetic field sensing by alternatively using pulsed (CPMG-like) or continuous (spinlock) dynamical decoupling techniques. We test the response of the $V_B^{-}$-based sensor to an oscillating magnetic field in the radiofrequency range of the form $B_{RF}(t)=b_{RF} $cos$(2\pi\nu_{RF} t + \phi)$, where $b_{RF}$ is the amplitude, $\nu_{RF}$ the frequency and $\phi$ the phase of the RF signal.

\begin{figure}[htbp]%
\centering
\includegraphics[width=1\textwidth]{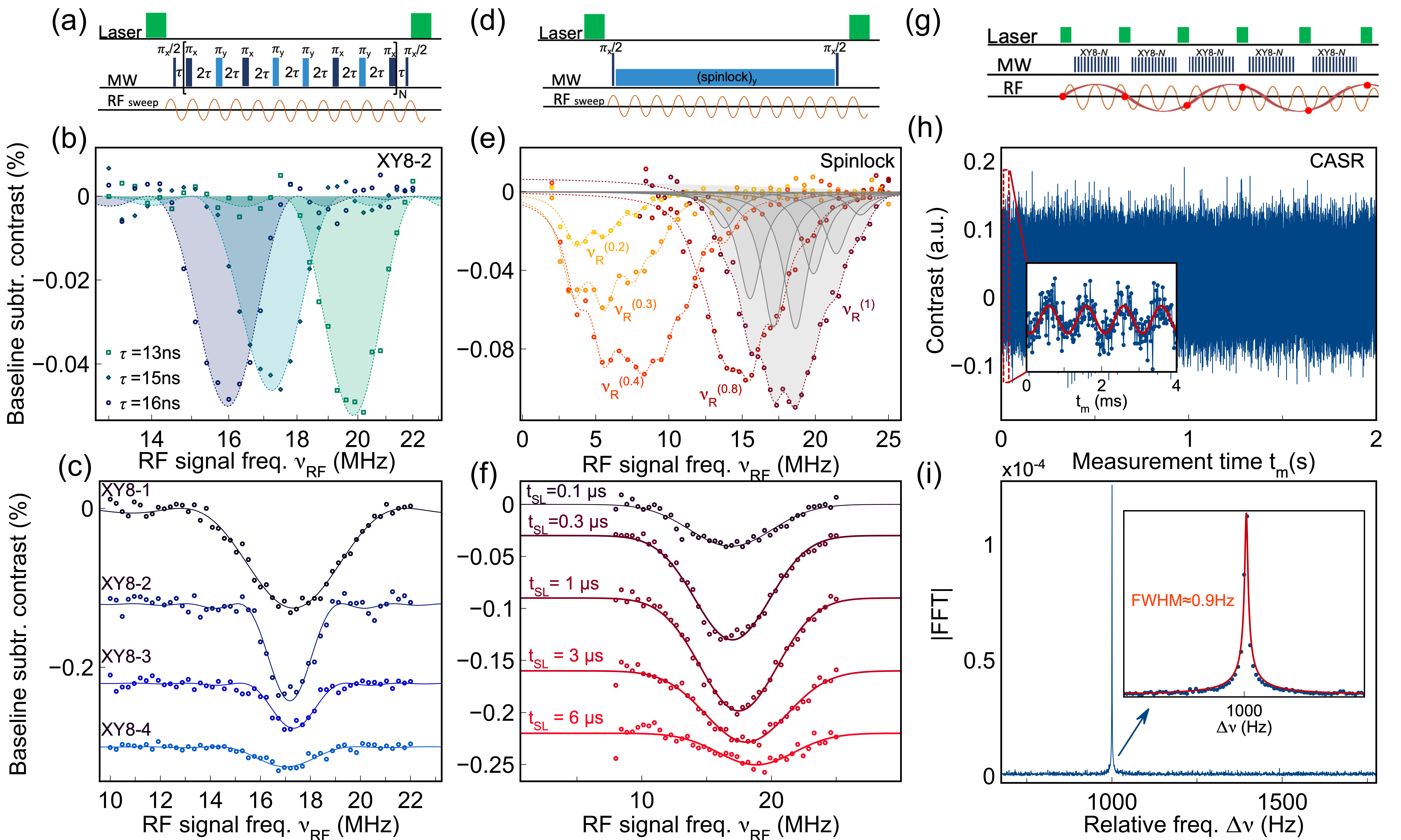}
\caption{\textbf{RF-sensing with $V_B^{-}$ in hBN.}
a) Pulsed-DD protocol used for sensing RF signals. An XY8-\textit{2} pulse sequence was utilized for the experiments in (b),(c) keeping $\tau$ fixed while sweeping the RF frequency.
b) Spectral response of the $V_B^{-}$ sensor as fluorescence contrast dips occurring at matching of the interpulse delay $\tau$ with the RF signal period according to: $\tau=1/(4\nu_{RF})$. The data points (dots) are best fitted with a (sinc)$^2$ function.
c) Dependence of the lineshape on the number of pulses for XY8-1,2,3,4 sequences setting $\tau=\SI{15}{\nano\second}$. 
d) Spinlock-based protocol for RF-sensing. The spinlock pulse duration was kept fixed ($t_{SL}=\SI{0.5}{\micro\second}$) and the RF frequency was swept. 
e) Spectral response as fluorescence contrast dips occurring at matching of the spinlock amplitude with the RF frequency according to $\nu_{R}=\nu_{RF}$. The labels next to each dip indicate decreasing values of $\nu_R$ which are normalized on the maximum spinlock amplitude used, corresponding to $\nu_R^{(1)}\sim\SI{18}{\mega\hertz}$.
f) Dependence of the lineshape on the spinlock duration used for the experiment with spinlock amplitude set to $\nu_{R}^{(1)}$.
g) Coherently-Averaged Syncronized Readout (CASR) protocol. XY8-\textit{2} subsequences are synchronized to the sensing frequency for 2 seconds.
h)Time-domain CASR signal. Each detected point corresponds to an optical readout represented by the red circles in (g). In the inset, a zoom-in of the signal in the first $\SI{4}{\milli\second}$ shows the oscillations at $\Delta\nu=\SI{1000}{\hertz}$. i)Fourier transformation of the signal in (h) yields a sharp peak at the relative frequency $\Delta\nu$. A Lorentzian linefit of the peak (red line) results in a $\SI{0.9}{\hertz}$ linewidth.}
\label{Fig3}
\end{figure}

As a first example, we use a pulsed dynamical decoupling sequence (XY8-\textit{N}), depicted in Figure \ref{Fig3}a. 
During the total free evolution time $t_s$, the oscillating magnetic field induces the accumulation of a relative phase $\theta$ to the $V_B^{-}$ coherence ($\vert\psi\rangle=(\vert0\rangle\pm e^{i\theta}\vert1\rangle)/\sqrt{2}$). The DD sequence then acts as a narrow-band RF filter and the $V_B^{-}$ superposition accumulates a maximal phase $\theta(t_s)=(2/\pi)\gamma b_{RF} t_s$\cite{abe_2016} if the condition $\tau=1/(4\nu_{RF})$ is matched, leading to a dip in the fluorescence intensity. In the experiment depicted in Figure \ref{Fig3}b, we study the sensor's spectral response by sweeping $\nu_{RF}$ while keeping $\tau$ at a defined value. Contrast dips appear at the expected RF frequencies for three different values of $\tau$. These dips can be fitted by the characteristic sinc-squared function\cite{degen_quantum_2017}(see Methods). 
In Figure \ref{Fig3}c, we show the dependence of the detected signal on the number of $\pi$-pulses. As anticipated, the XY8-\textit{N} sequence acts as a band-pass filter whose spectral width narrows down by increasing \textit{N}\cite{degen_quantum_2017}. However, simultaneously, the signal-to-noise ratio (SNR) decreases due to increasing decoherence. Furthermore, due to the short coherence time $T_2$, the pulsed DD sensing protocol fails at detecting frequencies
$\nu_{RF}\lesssim1/T_2\lesssim\SI{10}{\mega\hertz}$.

Figure \ref{Fig3}d displays an alternative approach for sensing RF signals based on rotating-frame magnetometry(\cite{loretz_radio-frequency_2013,hirose_continuous_2012,wang_nanoscale_2021}). It exploits the $V_B^{-}$ coherence locked on the transversal plane by a spinlock pulse of duration $t_{SL}$. 
Matching the $V_B^{-}$ Rabi frequency with the sensing frequency $\nu_R=\nu_{RF}$ drives transitions between $V_B^{-}$ dressed-states $\vert\pm\rangle=(\vert\ 0 \rangle\pm ie^{i\theta} \vert\ 1 \rangle)/\sqrt{2}$.
This has the consequence of inducing a Rabi evolution in the rotating frame 
that occurs with an oscillation rate set by the RF amplitude $b_{RF}$. This dressed-state dynamics determines the accumulation of a relative phase $\theta(t_{SL})=\frac{1}{2} \gamma b_{RF} t_{SL} $, required for sensing (see Supplementary Note 2). 
In Supplementary Note 3, we demonstrate coherent control of the $V_B^{-}$ dressed-states and probe their evolution during the spinlock pulse\cite{jeschke_coherent_1999,loretz_radio-frequency_2013,wang_nanoscale_2021}. 

By sweeping the frequency of the RF and performing the spinlock experiments with different MW amplitudes (see Supplementary Note 4), we observe dips similar to the data in Figure \ref{Fig3}b. However, the broad ODMR line and the MW field inhomogeneities throughout the $V_B^{-}$ ensemble lead to a distribution of effective Rabi frequencies as discussed in our previous work\cite{rizzato_polarization_2022} which result in broad dips that can be fitted as a sum of Lorentzian functions (Figure \ref{Fig3}b). 
We observe that the sensing  frequency bandwidth depends on the spinlock MW amplitude allowing access to frequencies $\nu_{RF}<\SI{10}{\mega\hertz}$ (see Figure \ref{Fig3}e). 
In addition, we study the spectral response of the sensor with different spinlock durations $t_{SL}$ (Figure \ref{Fig3}e). We observe a gradual increase of the SNR until an optimal duration of $0.5-1\SI{}{\micro\second}$ after which the signal starts to degrade, approaching the $T_1{\rho}$ and $T_1$ time limits. 
Line-broadening and a gradual shift to higher frequencies are also observed, likely due to heating caused by the long MW spinlock pulse. 
These results demonstrate how advanced quantum sensing protocols can be applied using spin defects in 2D materials and for qubit-control in 2D quantum registers\cite{liu_2d_2019,tadokoro_designs_2021,taminiau_2019,abobeih_one-second_2018,gao_nuclear_2022}.

\section{Sensing of RF signals with arbitrary frequency resolution}\label{sec5}
The frequency resolution in dynamical decoupling sequences is typically restricted by the duration of the pulse sequence and thus by the coherence time. This is in particular a limitation for $V_B^{-}$ defects in hBN, but it can be overcome by applying a sensing scheme called coherently averaged synchronized readout (CASR)\cite{glenn_high-resolution_2018,boss_quantum_2017,schmitt_submillihertz_2017}. 

Figure \ref{Fig3}g shows the pulse scheme, which consists of a train of DD sub-sequences (in our case XY8-\textit{2}), which are synchronized with the RF signal. 
If $\nu_{RF}\neq 1/4\tau$, consecutive phase shifts of the RF signal with respect to the DD sequences will occur, resulting in a downmixed fluorescence contrast oscillating at the difference-frequency $\Delta\nu=\nu_{DD}-\nu_{RF}$ ($\SI{1000}{\hertz}$ in our case). 
This scheme enables arbitrary frequency resolution since it is no longer limited by the intrinsinc coherence $T_2$ time of the solid-state spin system. 
We run the protocol for a total measurement time $t_m=\SI{2}{s}$ (Figure \ref{Fig3}h), resulting
in a time-dependent measurement that can be Fourier transformed to give a peak with sub-Hertz linewidth (Figure \ref{Fig3}h). 
Importantly, this method can render our ultra-thin quantum sensor capable of sensing oscillatory magnetic fields with a high-resolution, with possible applications in magnetic resonance spectroscopy\cite{allert_advances_2022}.
\section{Conclusion}\label{sec6}
This work addresses the short spin coherence of the hBN spin defects, a severe limitation for their application in quantum technology. 
Utilizing dynamical decoupling protocols, we achieve a $\sim50$ times extension of the coherence with respect to the single spin-echo $T_2$, approaching the limit of the longitudinal relaxation time $T_1$.
Furthermore, by generating $V_B^{-}$ dressed-states with spinlock pulses, we demonstrate an extension of the coherence up to $\sim\SI{7.5}{\micro\second}$, overcoming the spin-echo $T_2$ by nearly two orders of magnitude. 
The improved coherence times enable us to demonstrate RF signal detection in several complementary experiments.
In particular, we demonstrate that despite the intrinsically short $V_B^{-}$coherence, sensing RF frequencies with sub-Hz frequency resolution is possible.

The presented work paves the way to establish nanoscale magnetic resonance sensing and imaging using spin-based quantum sensors embedded in a van der Waals material.
Thanks to the possible miniaturization and integration in 2D heterostructures, these systems will render magnetic resonance techniques functional for exploring emergent phenomena in low-dimensional quantum materials and devices. 
Finally, the improved spin control and the intriguing nuclear-spin environment surrounding the defects open a promising path to the realization of multi-qubit registers for quantum sensors and quantum networks integrated into ultra-thin structures.
 
\clearpage
\section{Methods}\label{sec7}
\subsection{Experimental setup}
Initialization of the $V_B^{-}$ensemble is realized with a \SI{532}{\nano\meter} laser (Opus 532, Novanta photonics) at a power of $\sim\SI{100}{\milli\watt}$ (CW). Laser pulses are timed by an acousto-optic modulator (3250-220, Gooch and Housego) with typical pulse durations of \SI{5}{\micro\second}. The laser light is reflected by a dichroic mirror (DMLP650, Thorlabs) after which it is focused on the hBN flake by an objective (CFI Plan Apochromat VC 20x, NIKON) with a numerical aperture of NA=0.75.  Photoluminescence (PL) is collected by the same objective and alternatively focused by a tube lens on: 1) an Avalanche photodiode (APD) (A-Cube-S3000-10, Laser Components) for the spectroscopic path; 2) a camera (a2A3840-45ucBAS, Basler) for the imaging of the sample. The excitation green light and possible unwanted fluorescence from other defects are filtered out using a long-pass filter with a cut-on wavelength of 736 nm (Brightline 736/128, Semrock). The output voltage of the APD is digitized with a data acquisition unit (USB-6221 DAQ, National Instruments). An arbitrary waveform generator (AWG) up to 2.5GS/s (AT-AWG-GS2500, Active Technology) is used to syncronize the experiment and generate rectangular arbitrary-phased RF pulses for $V_B^{-}$ spin control. For up-conversion of the AWG RF frequency (typically 250 MHz) to the MW frequency required for $V_B^{-}$ driving, mixing with a local oscillator generator (SG384, Stanford Research Systems) is realized by means of an IQ mixer (MMIQ0218LXPC 2030, Marki). The amplified microwave pulses (ZHL-16W-43-S+, Mini-Circuits) are delivered by a gold strip-line to the hBN sample. A permanent magnet underneath the sample holder is utilized for applying the magnetic field of $\sim8$ mT. The ESR frequency is used to determine the magnetic field strength $B_0$ as well as the $V_B^{-}$\textsubscript{0,-1} resonance frequency \textit{f}\textsubscript{$V_B^{-}$}. The radiowaves signals used to demonstrate AC magnetometry are produced by an RF waveform generator (DG1022z, Rigol) connected to a 30 W amplifier (LZY-22+, Minicircuits). A small wire loop placed in the proximity of the sample was used for the RF delivery. 
\subsection{Sample preparation}\label{subsec2}
The hexagonal boron nitride van der Waals flakes were cleaved and tape-exfoliated starting from hBN seed crystals (2D semiconductors) on a silicon wafer with a $\SI{70}{\nano\meter}$ top oxide layer. We then implanted the exfoliated samples at the ion beam facility (Helmoltz-Zentrum Dresden-Rossdorf, HZDR) with a helium ion fluence of $10^{14}$ions/cm$^2$ at an energy of $3$ keV.
Once the photoluminescence spectrum of the boron vacancies was verified, we transferred the boron vacancy containing hexagonal boron nitride with a standard dry transfer method on top of a gold strip-line evaporated on a sapphire substrate. 
The gold strip-line was connected to a printed circuit board using several bond-wires in parallel to improve impedance matching and enabling high power microwave delivery for short Rabi pulses. 

\subsection{Pulse sequences, normalizations and fittings}\label{subsec3}
\subsubsection{Characterization of the $V_B^-$ spin defects}

\textbf{ODMR measurement.} For all experiments throughout this work, $\SI{5}{\micro\second}$-long laser pulses were used for initialization/readout. The ODMR measurement displayed in Figure \ref{Fig1}e) was performed using a $\SI{1}{\micro\second}$-long MW pulse at $\sim$ 1mW power for the driving of the $V_B^{-}$ spin-populations. The fluorescence contrast was monitored by increasing the MW frequency. For normalization and noise cancellation purposes, a second reference sequence was applied right after the first one, differing only from the MW being off\cite{bucher_2019}. The single data points result from dividing the fluorescence readouts of the two sequences. 10,000 averages for each data point and further 600 averages of the full frequency sweep were recorded. After baseline subtraction, the lineshape was fitted with seven Gaussian functions of the form $a[$exp$(-$ln$(2) (f-f_0)^2 / $LW$^2)]$ whose amplitude $a$ and frequency ($f_0$) were used as fitting parameters whereas the half-width-half-maximum LW was kept at a constant value of $\SI{22}{\mega\hertz}$. 

\textbf{Rabi nutation experiment.} Rabi oscillations were recorded by setting the MW frequency in the center of the ODMR spectrum and sweeping the MW pulse duration $t_p$ in stepwise increments. Different MW power (-5, -14 and -19dBm) were used corresponding to the different datasets in Figure \ref{Fig1}f). For each point, 10,000 averages were acquired following the same normalization and noise cancellation procedure used for the ODMR experiments. The Rabi oscillations were fitted using the following function: $1-c/2+c/2($cos$(2\pi\nu_R t+\Phi))[a\times $exp$(-b t)+m\times $exp$(-n t)]$ where $c$ is a term for coherence normalization, $\nu_R$ the Rabi frequency, $\Phi$ a phase term and the bi-exponential factor accounts for the damping of the oscillation.

$\mathbf{T_{1}}$ \textbf{measurement.} The $T_1$ time (see Figure \ref{Fig1}g) was measured by sweeping the time $\tau$ between initialization and readout pulses in a range from $\SI{10}{\nano\second}$ to $\sim\SI{30}{\micro\second}$. For noise cancellation purposes, a second reference sequence was applied right after the first one, differing only for a MW $\pi$-pulse inserted after the first laser pulse. The datapoints were then obtained by dividing the readouts of the two consecutive sequences. Every point was averaged 100,000 times and the whole time sweep averaged 4 times. The $T_1$ decay curve was fitted with a simple mono-exponential function of the form $a($exp$(-\tau/T_1))$ where $T_1=5.84\pm0.05 \si{\micro\second}$.

$\mathbf{T_{2}}$ \textbf{measurement.} A spin-echo sequence was set up following the scheme [$(\pi/2)_x$ - $\tau$ - $(\pi)_y$ - $\tau$ - $(\pi/2)_x$] where $\tau$ is swept from $\SI{10}{\nano\second}$ to $\sim\SI{200}{\nano\second}$.  Referencing for noise cancellation was achieved by alternating the last MW-pulse of the spin-echo sequence from $\pi/2$ to $3/2\pi$\cite{bucher_2019}. Every point was averaged 100,000 times and the whole time sweep averaged 5 times.
The decay was fitted with the function: $a($exp$(-2\tau/T_2)^c)+b($exp$(-d\tau))$cos$(2\pi f\tau)$ where the first stretched exponential yields the time constant $T_2=88.5\pm0.7 \si{\nano\second}$ with $c=1.37\pm0.01$. To guide the eye in the figures, we fitted the oscillation superimposed on the decay with a cosine function with $f=\SI{44.7}{\mega\hertz}$, multiplied by an exponential term which reproduces the damping of the observed oscillation with $d=1.2\times10^7$ s$^{-1}$.
\subsubsection{Extension of coherence}
\textbf{CPMG experiments.}
We applied the same scheme utilized for the $T_2$ measurement, that is monitoring the spin-echo signal while increasing the free evolution time $\tau$. Multiple decoherence curves were acquired for pulse sequences with growing number $N$ of $\pi$-pulses, up to $N=1000$. The data were normalized according to the following procedure. First, for all datasets the time axis was multiplied by a scaling factor $S=2N$ accounting for the real time elapsed between the first and the last $\pi/2$ pulse. Then, the decay curve corresponding to the single $\pi$-pulse experiment (spin-echo) was fitted by the function: $a($exp$(- t_s/T_2)^c)+b($exp$(-d t_s))$cos$(2\pi f t_s/S)$ (details see above). The point of maximum contrast, namely maximum intensity of the spin-echo, when no decoherence has occurred yet, has been extrapolated from the fit in correspondence to the time $t=0$ of the decay. Then, the contrast of all datasets has been normalized to this value and then fitted with the same function, keeping only the amplitudes $a$ and $b$ locked. The fitted $T_2$ values and relative exponents $c$ are reported in the table below.

\begin{table}[h]
\centering
    \caption{Fitted $T_2$ time constants and corresponding stretch-exponents $c$ for the data shown in Figures 2a,b.}
    \begin{tabular}{lll}
     \hline
    N ($\pi$-pulses)&$T_{2}$ [\si{\nano\second}]&$c$ \\
    \hline
1&$90\pm2$&$1.425\pm0.026$\\
4&$220\pm4$&$1.403\pm0.025$\\
16&$500\pm5$&$1.435\pm0.016$\\
64&$940\pm11$&$1.299\pm0.015$\\
300&$1759\pm32$&$1.309\pm0.024$\\
600&$2776\pm35$&$1.265\pm0.016$\\
800&$3434\pm30$&$1.257\pm0.011$\\
1000&$4204\pm44$&$1.310\pm0.014$\\
    \hline
    \end{tabular}
    \label{tab:TableS1.2}
\end{table}


$\mathbf{T_{1\rho}}$ \textbf{measurement.}
$T_{1\rho}$ was measured using a pulse sequence $[(\pi/2)_x - d - ($spinlock$)_y - d - (\pi/2)_x]$ and monitoring the resulting fluorescence contrast while applying step-by-step increments of the spinlock pulse duration $t_{SL}$. $\SI{3.5}{\nano\second}$-long $\pi/2$ pulses were used and the delay times $d$ kept to a value as short as possible ($\sim1-2$ ns), in order to avoid significant dephasing during this time. The spinlock amplitude was kept at 10\% of the MW amplitude utilized for the $\pi/2$-pulses. 100,000 averages were taken for each datapoint. The resulting curve fits a bi-exponential decay of the form: $a($exp$(- t_{SL}/T_{1a}))+b($exp$(- t_{SL}/T_{1b}))$, with $a=0.504\pm0.007$, $T_{1a}=1.38\pm0.03$ $\si{\micro\second}$, $b=0.554\pm0.003$, $T_{1b}=7.52\pm0.06$  $\si{\micro\second}$.

\subsubsection{AC-sensing}
\textbf{XY-\textit{N} protocol.}
AC field sensing was performed by means of a XY8-2 sequence following the scheme: $(\pi/2)_x[[-\tau-(\pi)_{\phi}-\tau]_8]_2$, with the following $\pi$ pulses phase-scheme: $\phi=[x-y-x-y-y-x-y-x]_2$. $\pi/2$ and $\pi$-pulse durations of respectively $\SI{3.5}{\nano\second}$ and $\SI{7}{\nano\second}$  have been used. The fluorescence contrast was monitored while the RF frequency $\nu_{RF}$ was swept in a range of a few MHz around the condition for the sensing frequency matching: $\nu_{RF}=1/4\tau$. Referencing for noise cancellation was achieved by alternating the last MW-pulse of the spin-echo sequence from $\pi/2$ to $3/2\pi$. Every point was averaged 100,000 times and the whole sweep averaged 25 times. The observed dips were first baseline-corrected by subtracting a line fitting the baseline in the spectral region without signal and then fitted with the following function: $a/2[$sin$(2\pi\tau N (\nu-\nu_{RF}))/2\pi\tau N(\nu-\nu_{RF})]^2$ leaving all 4 parameters $a$, $\tau$, $N$ and $\nu_{RF}$ free to fit. For instance, for the XY8-2 experiment set to $\tau=13$ ns,  $a=-1\times10^{-3}$, $\tau=0.013$, $N=16.1$ and $\nu_{RF}=19.8$. The same procedure has been applied for the data in Figure 3c.


\textbf{Spinlock protocol.}
For sensing, the same spinlock sequence described above for the measurement of the $T_{1\rho}$ was utilized. A $\pi/2$-pulse of $\SI{3.5}{\nano\second}$ and a spinlock duration $t_{SL}=\SI{0.5}{\micro\second}$ were used. The fluorescence contrast was monitored while the RF frequency $\nu_{RF}$ was swept in a range of a few MHz around the matching condition: $\nu_R=\nu_{RF}$. Five different values of Rabi frequency were used, setting the spinlock MW to the 2\%, 3\%, 5\%, 9\%, 11\% of the MW amplitude utilized for the $\pi/2$-pulses. The Rabi oscillations corresponding to the latter two spinlock amplitudes were recorded for a check (Supplementary Note 4). The observed dips were first baseline-corrected by subtracting a line fitting the spectral region without signal and then fitted with several Lorentzian functions: $a/(1 + (\nu-\nu_{RF})^2 / $LW$^2)$ whose sum reproduces the overall lineshape. In these fits only the linewidth of the Lorentzians was kept locked to $1/t_{SL}=\SI{2}{\mega\hertz}$. Same pulse sequence and same background subtraction was applied in Figure 3f. In this case the fitting was performed by simple Gaussian functions. 

 \textbf{CASR protocol.}
We applied a pulse sequence which was synchronized with the sensing RF signal and consisted of concatenated XY8-\textit{2} sub-sequences repeated so that the timing between them was an integer number of the RF period.
For each XY8 sequence, we used $\SI{5}{\micro\second}$ long laser pulses and $\pi$-pulses of $\SI{8}{\nano\second}$ for initialization/readout and MW manipulation, respectively. We sensed an RF frequency $\nu_{RF}\sim\SI{18}{\mega\hertz}$ by setting the interpulse delay $\tau=\SI{14}{\nano\second}$.   To get a relative frequency $\Delta\nu=\SI{1000}{\hertz}$, the RF was shifted to $\nu_{RF}=\SI{18.001}{\mega\hertz}$. 
A total measurement time $t_m$ of 2 seconds was utilized. The detected time trace was then Fourier transformed and the magnitude plotted in Figure 3i. The resulting peak in the frequency domain was fitted with a modified Lorentzian function to determine the full width half-maximum (FWHM). The signal was the result of 1000 averages.

\newpage
\section{Figures}\label{sec8}

\clearpage
\section*{Acknowledgment} 
R.R. and D.B.B. would like to thank Prof. Dr. Steffen Glaser for stimulating discussions. R.R. thanks Nick Neuling and Giovanna Salvitti for their help in the lab. This study was funded by the Deutsche Forschungsgemeinschaft (DFG, German Research Foundation) - 412351169 within the Emmy Noether program. R.R. acknowledges support from the DFG Walter Benjamin Programme (Project RI 3319/1-1). J.J.F. and D.B.B. acknowledges support from the DFG under Germany’s Excellence Strategy—EXC 2089/1—390776260 and the EXC-2111 390814868. Support by the Ion Beam Center (IBC) at HZDR is gratefully acknowledged.

\section*{Additional Information}
\textbf{Author Contributions}
R.R. M.S. and D.B.B. conceived the idea of AC sensing with $V_B^-$ in hBN. R.R. and D.B.B. designed the research. R.R. carried out the experiments. M.S. and P.J. prepared the sample and fabricated the microstructure for MW delivery. R.R. M.S. and S.M. built the ODMR setup. F.B. programmed the pulse sequences, J.C.H. and J.P.L. helped in the optimization of the experimental setup, J.C.H. contributed with the theoretical derivations. G.V.A., U.K. and M.H. were responsible for the ion implantation for generation of the $V_B^-$ centers in the hBN substrate. A.V.S. and J.J.F. advised on several aspects of theory and experiments. R.R., M.S. and D.B.B. analyzed the data. R.R. and D.B.B. wrote the manuscript with inputs from all authors. All authors reviewed the manuscript and suggested improvements. 
All correspondence and request for materials should be addressed to R.R.(roberto.rizzato@tum.de) or D.B.B. (dominik.bucher@tum.de).

\textbf{Competing interests}. All authors declare that they have no competing interests. 

\textbf{Data and materials availability}. All data needed to evaluate the conclusions in the paper and/or the Supplementary Materials. Additional data related to this paper may be requested from the authors. 


\clearpage
\bibliography{bibliography}
\clearpage
\section*{Supplementary Information}
\renewcommand{\thefigure}{S\arabic{figure}}
\setcounter{figure}{0}
\renewcommand{\thetable}{S\arabic{table}}
\setcounter{table}{0}
\subsection*{Supplementary Note 1: ESEEM Modulation}
In the $T_2$ measurement reported in Figure 1h and in the CPMG experiments in Figure 2a, we observe a strong modulation of the time-domain electron spin signal reminiscent of the electron spin envelope modulation (ESEEM) observed in other systems when the central electronic spin strongly interacts with proximate nuclei, such as \textsuperscript{14}N and \textsuperscript{13}C. Fitting this oscillation with a function $F(t)\propto $cos$(2\pi f t)$ yields a frequency of $f\sim\SI{45}{\mega\hertz}$, matching the hyperfine coupling of the three equivalent \textsuperscript{14}N surrounding the defect. A similar effect can also be predicted by using a well-known EPR calculation toolbox\cite{stoll_easyspin_2006} and simulating a two-pulse ESEEM experiment using the following parameters: spin S=1;g-factor=2;zero-field-splitting: D=3.47 GHz,E=60 MHz. Three \textsuperscript{14}N have been considered in the spin system, coupled to the electronic spin with the following hyperfine tensors A1=[80 57 47], A2=[46 91 48], A3=[80 57 47]\cite{ivady_ab_2020,liu_coherent_2022}. Further parameters: ext.field = 8 mT, MWfreq=3.2 GHz, excit. bandwidth=250 MHz, Exp.Sequence=2pESEEM, resolution dt=0.001$\mu$s. Relaxation parameters: $T_1$=6$\mu$s, $T_2$=80 ns.
\begin{figure}[h]%
\centering
\includegraphics[width=1\textwidth]{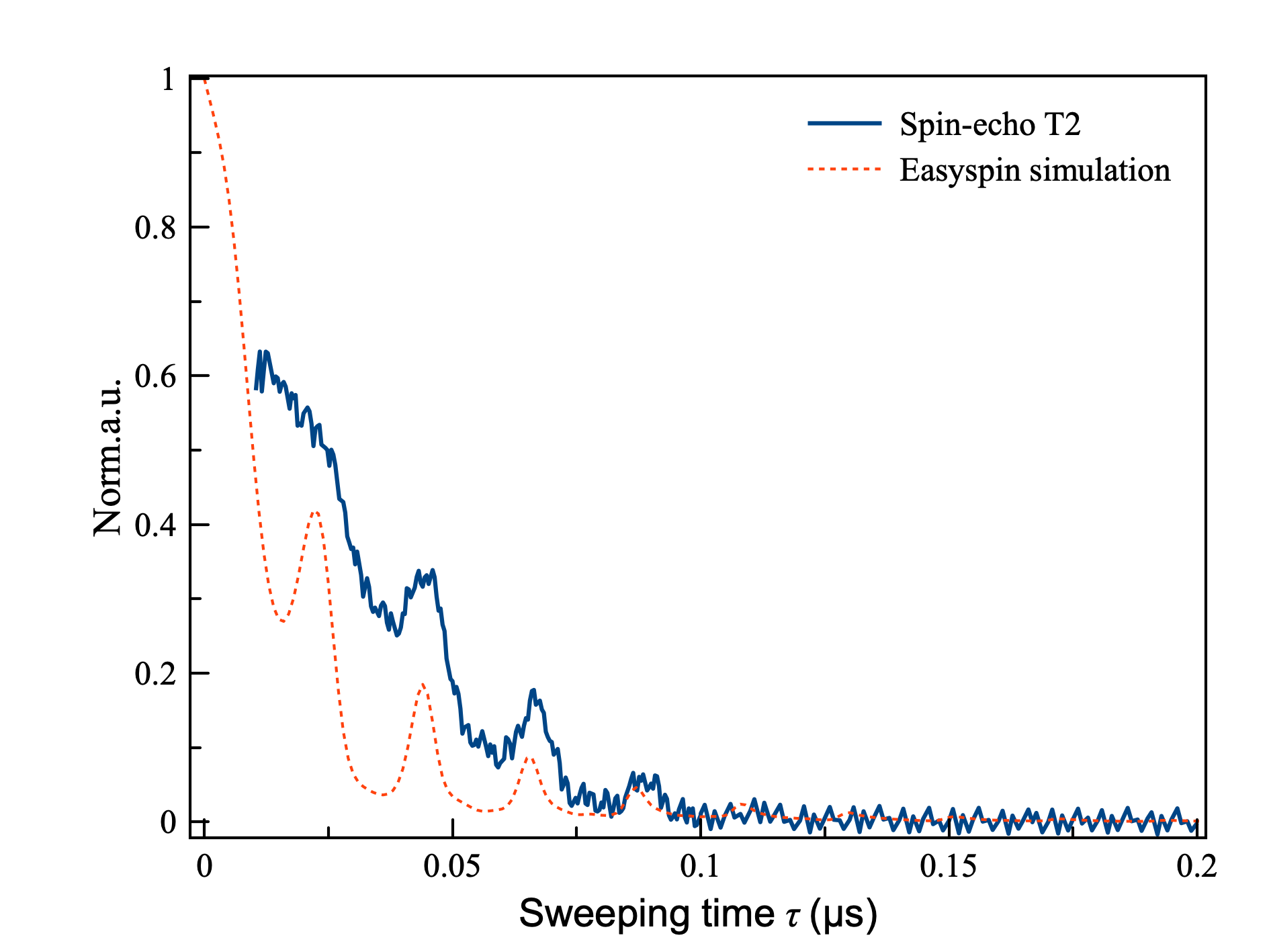}
\caption{\textbf{Simulation of a two-pulse ESEEM experiment.}
 A two-pulse ESEEM experiment is simulated with the Easyspin function "saffron".}\label{FigS1}
\end{figure}

\clearpage
\subsection*{Supplementary Note 2: Phase accumulation in the spinlock experiment}
The rotating-frame Hamiltonian $H_\text{rot}$, accounting for a spinlock pulse of amplitude $\Omega$ around the $y$-axis and an AC magnetic field signal $B_{RF}(t)=b_{RF}$cos$(2\pi\nu_{RF} t + \phi)$ along the $z$-axis, reads (in natural units $\hbar=1$):
\begin{equation}
    H_\text{rot}=\frac{1}{2} \Omega \, \sigma_y+\frac{1}{2} \gamma b_{RF} \cos(2\pi\nu_{RF} t + \phi) \, \sigma_z \,,
\end{equation}
 with the Pauli matrices $\sigma_y$ and $\sigma_z$. This equation can be transferred into a second rotating frame of the Rabi oscillation, where the first term vanishes. Assuming that the matching condition of $\Omega=2\pi \nu_{RF}$ is fulfilled and that the rotating wave approximation for $2\Omega  \gg b_{RF}$ is valid, the Hamiltonian $H_\text{rot,2}$ in the second rotating frame becomes:
\begin{equation}
   H_\text{rot,2}= \frac{1}{4} \gamma b_{RF} \left( \cos (\phi) \sigma_z + \sin (\phi) \sigma_x \right) \, .
\end{equation}

Setting the signal phase to $\phi=0$ and implementing the Hamiltonian in a time evolution operator $U(t)$ leads to the following:
\begin{equation}
    U(t)=\exp(-\mathrm{i}Ht)=\exp\left( -\frac{\mathrm{i}}{4} \gamma b_{RF}t \, \sigma_z \right) \, .
    \label{eq:propagator}
\end{equation}
This equation can be connected to the general expression for a rotation operator
\begin{equation}
    R_\theta=\exp\left( -\frac{\mathrm{i}}{2}  \theta \sigma_z  \right) \, ,
\label{eq:rot-operator}
\end{equation}
which describes a rotation of the Bloch vector by an angle of $\theta$ around the $z$-axis. Comparing equations \eqref{eq:propagator} and \eqref{eq:rot-operator}, gives the phase $\theta$ that is accumulated during the spinlock pulse on the equatorial plane in the second rotating frame:
\begin{equation}
    \theta(t)=\frac{1}{2} \gamma b_{RF} t \, .
\end{equation}

\clearpage
\subsection*{Supplementary Note 3: Coherent manipulation of the $V_B^-$ dressed-states}
\begin{figure}[h]%
\centering
\includegraphics[width=0.85\textwidth]{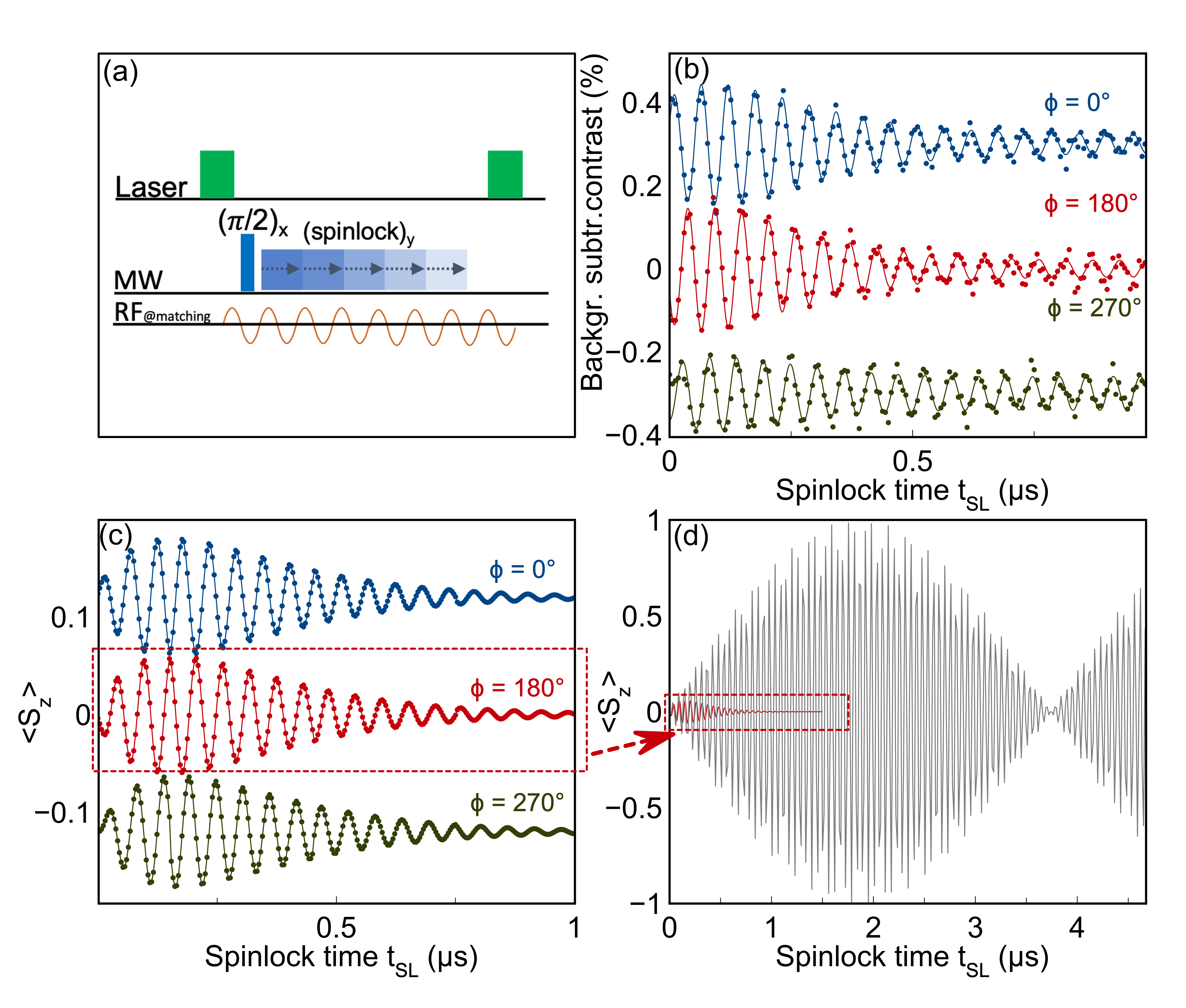}
\caption{\textbf{Dressed-state Rabi oscillations}
a) Spinlock pulse sequence. Each datapoint correspond to the fluorescence contrast detected immediately after the spinlock pulse. b) Experiments performed setting different $\phi$ values resulting in different Rabi phases. c)Numerical simulations of dressed-state Rabi nutation. d)Overlap of the simulation reported in c) (red line) and the same numerical simulation with $t_{SL}\sim\SI{5}{\micro\second}$ in the absence of decoherence (grey). 
} \label{FigS3}
\end{figure}

In the following experiments the evolution of the $V_B^{-}$ dressed-states during matched spinlock is probed. Here, a spinlock pulse sequence is applied where the spinlock duration is gradually increased. The RF frequency is kept at a fixed value, matching the condition $\nu_{RF}=\Omega/{2\pi=18}$ MHz. The RF phase $\phi$ is kept constant with respect to the pulse sequence. Furthermore, the last $\pi/2$-pulse has been removed (see Figure \ref{FigS3}a) to directly detect the rotating-frame Rabi evolution as oscillations of the spin vector's $z$-projection. 
The observed signal is modulated at a frequency of $\SI{18}{\mega\hertz}$ in accordance to the matching condition. Moreover, the phase of these oscillations is sensitive to the RF phase $\phi$ utilized for driving the dressed-states (see Figure \ref{FigS3}b). To, at least qualitatively, describe the dynamics occurring during the experiment, numerical simulations have been performed based on the density matrix evolution of a simple spin-1/2 system under the Hamiltonian reported in the equation 1 of the Supplementary Note 2. The resulting density matrix is then corrected by an exponential term accounting for dressed-state decoherence using a bi-exponential decay with time constants of $T_a=1.6\times10^{-7}$ s and $T_b=1.5\times10^{-6}$ s with amplitudes of 0.95 and 0.05, respectively. These parameters have been chosen until the oscillation damping in the simulation resembled the experimental one. The simulation for $\phi=180\deg$ with no relaxation correction is shown in (Figure \ref{FigS3}d) for a comparison. We note that, due to the RF field amplitude not being strong enough to counteract the fast decoherence, the detected evolution is likely to be only the very initial part of the complete Rabi rotation from the $\vert+\rangle$ to $\vert-\rangle$ dressed-states (Figure \ref{FigS3}d).

\clearpage
\subsection*{Supplementary Note 4: Measurement of Rabi frequencies for spinlock matching}
Spinlock sensing is accomplished by setting the spinlock MW amplitude such that the corresponding $V_B^-$ Rabi frequency matches the RF sensing frequency. In this work, this is practically done by setting the AWG amplitude to a fraction of the available range, only for the spinlock pulse. For a check, Rabi experiments have been performed by setting, as the Rabi-pulse amplitude, the same used for the spinlock pulse. Here, Rabi oscillations corresponding to the experiments labeled $\nu_R^{0.8}$ and $\nu_R^{1}$ are reported. As visible, they yield Rabi frequencies respectively of 15 and 18 MHz, exactly where the dips fall in Figure 3e.
\begin{figure}[h]%
\centering
\includegraphics[width=0.8\textwidth]{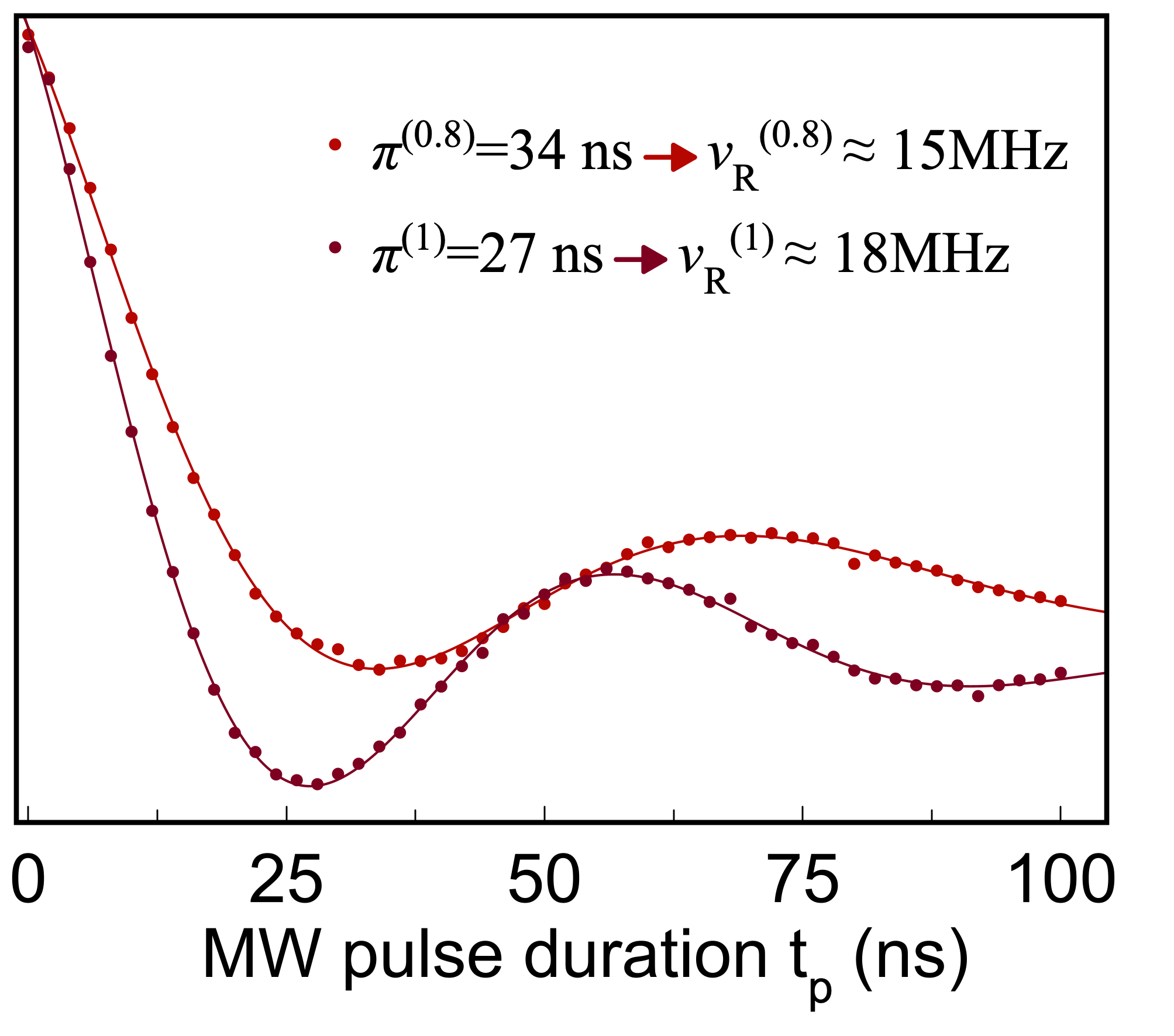}
\caption{\textbf{Measurement of Rabi frequencies corresponding to spinlock MW amplitudes. }
}\label{FigS2}
\end{figure}
\clearpage
\subsection*{Supplementary Note 5: Spinlock sensing of RF fields by sweeping the spinlock amplitude}
In the main text, the proof-of-principle demonstration of RF sensing by the spinlock protocol (rotating frame magnetometry) was accomplished by setting the MW spinlock amplitude at a defined value and sweeping the RF signal frequency $\nu_{RF}$ until the matching condition for phase accumulation was met and a dip in the fluorescence contrast was observed. 
Here, we show the vice-versa experiment, where the spinlock amplitude $A_{SL}$ is swept until matching a certain RF frequency. Although the same fundamental physics is involved, this sensing mode demonstrates the capability of the sensor to seek an unknown RF signal frequency.  

\begin{figure}[h]%
\centering
\includegraphics[width=0.8\textwidth]{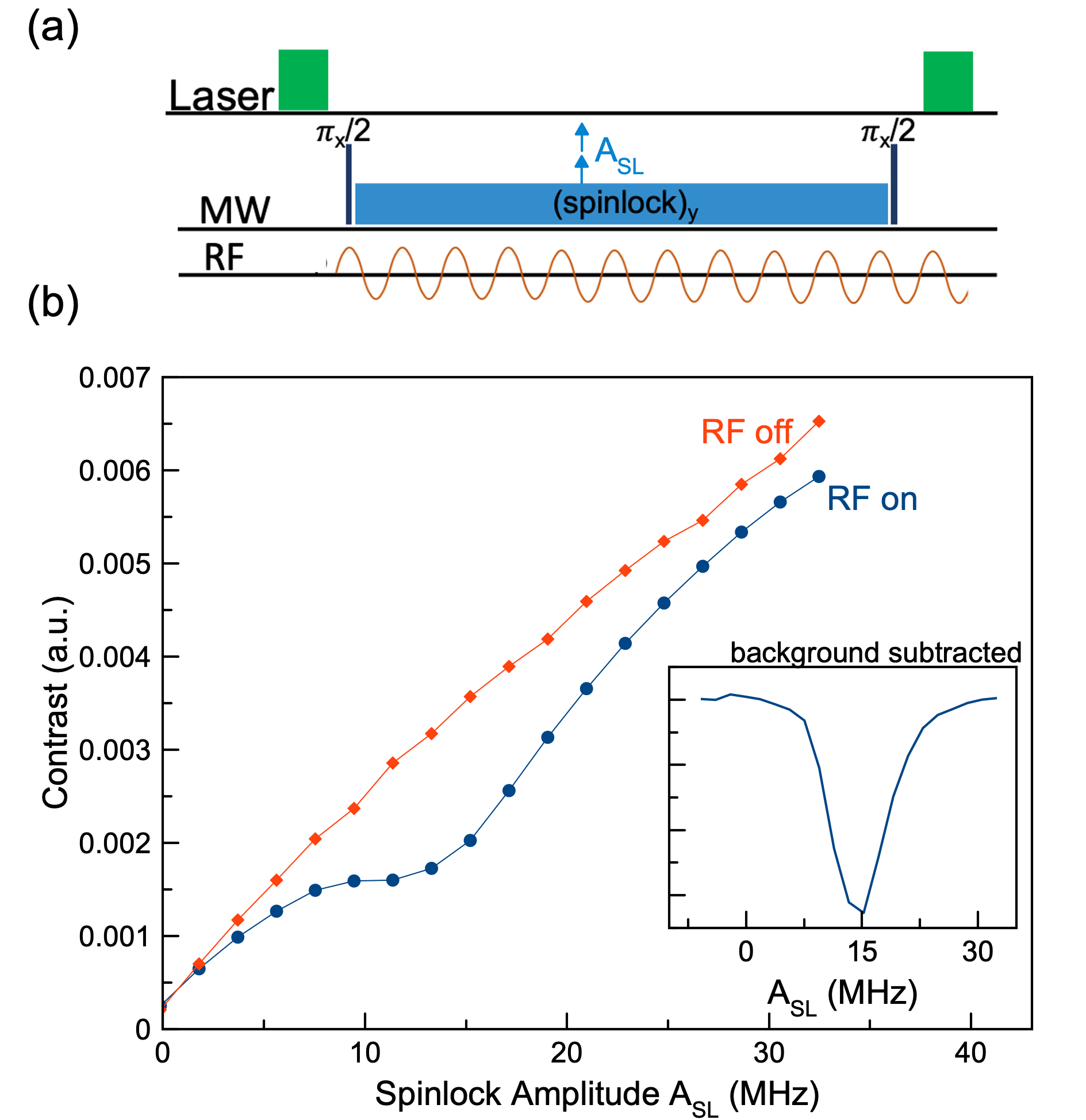}
\caption{\textbf{Spinlock sensing by spinlock amplitude sweep.} a) Pulse sequence utilized for the experiment. The RF frequency was kept fixed at a value of $\nu_{RF}=\SI{14}{\mega\hertz}$ and the spinlock MW amplitude $A_{SL}$ was swept in a range around $\nu_{RF}$. b) Spinlock amplitude sweep experiment keeping the RF signal off (red line) and on (blue line). A broad dip appears centered at a spinlock amplitude value matching the RF frequency. In the inset the dip is shown after baseline subtraction. The spinlock amplitude, originally in units of Volt-per-peak (Vpp) was converted to MHz by performing a calibration matching several RF frequencies at different values of $A_{SL}$, as done in \cite{rizzato_polarization_2022}.
}\label{FigS4}
\end{figure}
\end{document}